\documentclass[12pt]{article}
\usepackage{amsfonts,epsfig}

\def\pa{\partial}

\def\k{\kappa}
\def\d{\delta}
\def\h{\hat}
\def\b{\bar}
\def\t{\tilde}
\def\f{\frac}

\def\p{\varphi}
\def\k{\kappa}
\def\d{\delta}
\def\h{\hat}
\def\t{\tilde}
\def\f{\frac}
\def\dt{\dot}

\def\l{\label}
\def\e{\varepsilon}
\def\a{\alpha}
\def\bt{\beta}
\def\la{\lambda}
\def\g{\gamma}

\def\m{\mu}
\def\n{\nu}

\def\r{\rho}

\def\s{\sigma}

\def\th{\theta}

\def\be{\begin{equation}}
\def\ee{\end{equation}}
\def\ba{\begin{eqnarray}}
\def\ea{\end{eqnarray}}
\def\z{\bar{z}}
\def\c{\chi}

\begin{document}

\vspace{4cm}
\centerline{\large\bf Near-horizon symmetries of the Schwarzschild
 black holes with supertranslation field}
\vspace{1cm}
\centerline{\bf Mikhail Z. Iofa \footnote{ e-mail:
iofa@theory.sinp.msu.ru}}
\centerline{Skobeltsyn Institute of Nuclear Physics}
\centerline{Moscow State University}
\centerline{Moscow 119992, Russia}

\begin{abstract}
Using the exact solution to Einstein equations of Compere and Long
for the Schwarzschild metric
containing  supertranslation field,
we study the near-horizon symmetries
of the metric. We consider class of metrics with supertranslation field
depending only on a spherical angle $\theta $. After reviewing the action of
supertranslations preserving the static gauge of the metric, we
study general transformations preserving the near-horizon form of the metric.
 We determine the asymptotic Killing vectors and
 calculate the charge corresponding to the asymptotic symmetries
preserving the form of the metric at the horizon.
\end{abstract}

\section{Introduction}
 
The study of asymptotically flat spacetimes initiated by Bondi, Metzner and 
Sachs (BMS) \cite{bondi,sachs} has attracted much attention recently
(for a review and references \cite{strom1}). The symmetry group 
of the asymptotically flat gravity extends the Poincare group and contains 
supertranslations, the angular-dependent translations at null infinity,
which generalize translations. 
The vacua with different supertranslation fields are physically different
in a sense that their (superrotation) charges are different 
\cite{bar3,bar2,bar4,charge,comp1,comp2}.
Different vacua differ one from  another by the cloud of massless particles 
they contain \cite{fad,strom4}. 

Of particular interest is the study of black holes containing supertranslation
field. It was suggested that evaporation of black holes with inclusion of
 soft quanta may be unitary \cite{fad,carney,strom4}.
Finite supertranslation diffeomorphisms map physical states to inequivalent 
physical states having different charges.
Stationary black holes resulting from the collapse 
are diffeomorphic to the Kerr metric \cite{carter,robinson} and have the same ADM
charges. Because diffeomorphisms contain supertranslations, the
 metric of a black
hole, in general, contains  supertranslation field.
 It was discussed the possibility that the account of
supertranslation (and superrotation) hair may contribute to solving the information
problem \cite{hawk3,hawk31,bousso1,bousso2}. 

The BMS transformations are naturally formulated at the null infinity, and there is
a complicated problem of extension of asymptotically defined metric with 
supertranslation field in a closed form to the bulk. In paper \cite{comp1} a family 
of vacua 
containing supertranslation field was constructed in the bulk. In paper \cite{comp2} a
solution-generated technique was developed and applied to construction of solutions
to the Einstein equations containing  supertranslation field   
generalizing the Schwarzschild metric. 

Because many properties of black holes are associated with physics near horizon, and
having at hand the Schwarzschild metric containing supertranslation field, it is natural
to study its near-horizon symmetries.  

We consider the class of metrics containing supertranslation field
depending only on a spherical angle $\th$.
The metrics are solutions of the Einstein equations and are generalization of
the Schwarzschild solution.
We transform the metric to a coordinate system in which horizon of the metric is located
at $r=2M$.
A specific property of this  metric is that,
in distinction to the metrics without supertranslation field, the expansion
of the metric near horizon is in powers of $(r-2M)^{1/2}$ and not in powers
of $r-2M$ \cite{booth,wald}.

 We review the action of supertranslations preserving
the gauge of the metric \cite{comp2}.
We find sufficient condition on supertranslations $T(\th )$ preserving
the gauge of the metric to preserve the near-horizon form of the metric.

We perturbatively solve equations for the asymptotic Killing vectors preserving 
the functional form of the metric at the horizon 
(cf.\cite{carlip,koga,hotta,padm,donnay41,donnay4,akh,set} and many other papers).
  We calculate the variation of the 
surface charge corresponding to the asymptotic horizon symmetries. 
We find that in the case
of the black hole with  supertranslation field $C(\th )$ depending only on 
spherical angle $\th$ the variation can be integrated
to a charge in a closed form.

\section{Schwarzshild metric containing supertranslation field}

In this section we briefly review the
 exact solution of the Einstein equations
generalizing the Schwarzschild metric and containing supertranslation field
constructed in paper \cite{comp2}. Next, we transform the metric of \cite{comp2}
to a form with the horizon located at the surface $r=2M$, where $M$ is interpreted as
a mass of black hole. 

Vacuum solution of the Einstein equations containing supertranslation field 
diffeomorphic to the Schwarzschild metric is
\ba
\l{2.1}
&{}&ds^2 =\t{g}_{mn}dx^m dx^n=\\\nonumber
&{}&=-\f{(1-M/2\r_s )^2}{(1+M/2\r_s )^2}dt^2 +
(1+M/2\r_s )^4 \left(d\r^2 + (((\r -E)^2 +U)\g_{AB} +(\r -E)C_{AB})dz^A dz^B
 \right).
\ea
$C(\th ,\p )$ is a supertranslation field, $C_{00}$ is the lowest spherical harmonic 
of $C(\th ,\p )$. In the following we do not write $C_{00}$ explicitly understanding
$C\rightarrow C-C_{00}$.
Here 
\be
\l{2.2}
\r_s (\r, C) =\sqrt{(\r -C )^2 + D_A C D^A C}
.\ee
Covariant derivatives $D_A $ are  defined
with respect to the metric on the sphere
$ds^2 =d\th^2 + \sin^2 \th d\p^2$.
The tensor $C_{AB}$ and the functions $U$ and $E$ are
\ba
\l{2.3}
&{}& C_{AB}=-(2D_A D_B -\g_{AB} D^2 ) C, \\\nonumber
&{}& U=\f{1}{8}C_{AB}C^{AB},\\\nonumber
&{}& E=\f{1}{2}D^2 C +C.
\ea
In the following we consider the case of  metrics with $C$ depending
only on angle $\th$. In this case the components $C_{AB}$ are
\ba
\l{2.4}
&{}&C_{\th\th}=-(C''-C'\cot\th ),\\\nonumber
&{}&C_{\p\p}=\sin^2 \th (C''-C'\cot\th ),\\\nonumber
&{}&C_{\th\p}=0.
\ea
Here prime is differentiation over $\th$.
Horizon of the metric (\ref{2.1}) is located at the surface ( see\cite{comp2}, Eq.(45))
\be
\l{h}
\r_H =C +\sqrt{\f{M^2}{4} -D_A C D^A C}
.\ee
To obtain another form of the metric with horizon located at $r=2M$,
we introduce a new variable $r= r(\th, \r )$ defining
\be
\l{2.5}
r=\r_s (\r, C)\left(1+\f{M}{2\r_s (\r, C)}\right)^2
.\ee
The inverse transformation $\r =\r(\th ,r)$ is obtained from the relation
\be
\l{2.6}
\sqrt{(\r -C )^2 + {C'}^2}=\f{1}{2}(r-M+\sqrt{r(r-2M)}).
\ee
Introducing the functions
\ba
\l{2.7}
&{}&V=1-\f{2M}{r},\\
\l{2.8}
&{}&K=r-M+\sqrt{r(r-2M)},\qquad \f{dK}{dr}=\f{K}{rV^{1/2}}
\ea
we have
\be
\l{2.9}
\f{(1-M/2\r_s )^2}{(1+M/2\r_s )^2}=V,\qquad (1+M/2\r_s )^4 =\f{4r^2}{K^2}
.\ee
Expressing  $\r$ as a function of $r$ and $C$, we obtain
\be
\l{2.10}
\r =C +\f{K}{2}\left(1-\f{4{C'}^2}{K^2}\right)^{1/2}.
\ee
Denoting 
\be
\l{2.11}
b =\f{2C'}{K}
,\ee
we have
\be
\l{2.11a}
d\r =\f{K}{2}\left[\left(b-\f{bb'}{\sqrt{1-b^2}}\right)d\th
+\f{dr}{rV^{1/2}\sqrt{1-b^2}}\right]
.\ee
Because $K(r)$ is an increasing function of $r$ having its minimum at $r=2M$, 
and because the cosmic censorship
conjecture implies (see  Eq.(47) of \cite{comp2}) that
$$
1-\f{{C'}^2}{M^2}>0
,$$
we have $1-b^2 >0$.
We obtain the line element (\ref{2.1}) in a form \cite{iofa}
\ba
\l{2.12}
&{}&ds^2 =g_{\m\n} dx^\m dx^\n
 =-Vdt^2 +  \f{dr^2}{V(1-b^2 )} +2drd\th\f{br (\sqrt{1-b^2}-b' )}{(1-b^2 )V^{1/2}}
+\\\nonumber
&{}& + d\th^2 r^2\f{ (\sqrt{1-b^2}-b' )^2}{(1-b^2 )} +d\p^2 r^2\sin^2\th (b\cot\th
-\sqrt{1-b^2 })^2 =\\\l{2.13}
&{}&=-Vdt^2 +dr^2 \,\f{\b{g}_{rr}}{V} +2dr d\th\,\f{\b{g}_{r\th}}{V^{1/2}}+
d\th^2 \,\b{g}_{\th\th}+ d\p^2 \sin^2\th\,\b{g}_{\p\p}
\ea
Horizon of the metric (\ref{2.12}) is located at the surface $r_H =2M$.
 In Appendix A we consider solution of the geodesic equations
for null geodesics in the metric (\ref{2.12}). We find the asymptotic of the solution
in the limit $V(r)\rightarrow 0 $. In this limit 
$$
\f{dr}{dt}= const V(r)
,$$
i.e. the same relation as in the Schwarzschild metric. The surface $r=2M$ 
is the surface of the infinite redshift, i.e. horizon \cite{LL}.
In the following, we set $M=1$.

\section{Supertranslations preserving the gauge of the metric}

In this section, first, we review the supertranslations preserving 
the gauge of the metric
(\ref{2.1}). Next, we write the generator of supertranslations preserving the gauge of
the metric (\ref{2.12}). Imposing the requirement that supertranslations do not
alter the functional form of the metric at the horizon, we determine
necessary conditions on supertranslations to preserve the near-horizon form of
the metric. We begin with the metric component $g_{tt}$ and
find conditions on supertranslations under which the functional form of 
 $g_{tt}$ at the horizon is preserved. 
Considering other components of the metric, we show that 
the condition obtained for  $g_{tt}$
is sufficient to preserve the near-horizon form of all components.
  
The metric (\ref{2.1}) is written in the static gauge $\t{g}_{t\r}=\t{g}_{t\th}=0$ and
$\t{g}_{\r a}=0$.
Generator of supertranslations preserving this gauge was obtained in \cite{comp2}, and
in the case $C=C(\th )$ is
\be
\l{3.1}
\xi_T =T_{00}\f{\pa}{\pa t}-(T-T_{00})\f{\pa}{\pa \r}-\f{T'}{\r -C -C''}\f{\pa}{\pa \th}
.\ee
In the following we include $T_{00}$ in $T$ writing simply $T$.
The metric components transform as
\be
\l{3.1a}
L_{\xi_T}\t{g}_{\m\n}={}{\lim}_{\e =0}\left[\t{g}_{\m\n}(C+\e T)-
\t{g}_{\m\n}(C)\right]/\e \equiv \d_T \t{g}_{\m\n}(C)
,\ee
where $\d_T C=T$ and $\d_T D^k C=D^k T$.
The action of the transformation (\ref{3.1}) on  the component $\t{g}_{tt}$  is
\be
\l{3.3}
L_{\xi_T}\t{g}_{tt} = 4\f{\r_s -1/2}{(\r_s +1/2)^3}\d_T \r_s ,
\ee
where
\be
\l{3.2}
\d_T \r_s =\f{-T(\r -C )+C' T'}{\r_s}
.\ee
Horizon of the metric (\ref{2.1}) is located at the surface $\r_s =1/2$.

Transformations (\ref{3.1})  form a  commutative algebra w.r.t. the modified
bracket \cite{bar3,bar2,comp2}
\be
\l{3.12}
[\xi_1,\xi_2]_{mod}=[\xi_1,\xi_2] -\d_{T_1}\xi_2 +\d_{T_2}\xi_1
.\ee
Because $[\xi_1,\xi_2]^\r = 0$ and $\d_{T_1}\xi_2^\r=0$,
the commutator $[\xi_1,\xi_2]_{mod}^\r$ is zero. The commutator $[\xi_1,\xi_2]^\th$
is equal to
\be
\l{3.13}
[\xi_1,\xi_2]^\th =\f{T_{12} +T'_{12}}{(\r-C-C'')^2}
,\ee
where $T_{12}=T_2 T'_1 -T_1 T'_2$, and
\be
\l{3.14}
\d_{T_1}\xi^\th_2 =-T'_2 \f{T_1 + T''_1 }{(\r-C-C'' )^2}
\ee
giving for  $[\xi_1,\xi_2]_{mod}^\th $ zero result.

Next, we consider supertranslations acting on the metric (\ref{2.12}).
The generator of supertranslations in variables $(r, \th )$ acting on
 the metric (\ref{2.12}) is obtained from the generator
(\ref{3.1}) by transformation
\ba
\l{3.6}
&{}&\nonumber
\c^r_T =\xi^\r \f{\pa r}{\pa\r}+\xi^\th \f{\pa r}{\pa\th}=
\xi^\r \f{\pa r}{\pa\r_s}\f{\pa \r_s}{\pa\r} +
\xi^\th \f{\pa r}{\pa\r_s}\f{\pa \r_s}{\pa\th}\\
&{}&\c^\th_T =\xi^\r \f{\pa \th}{\pa\r}+\xi^\th \f{\pa \th}{\pa\th}=\xi^\th_T.
\ea
Using (\ref{2.5}) to calculate $\pa r/\pa\r_s =1-1/(4\r_s^2 )$,
 and expressing $\c_T^i$ through $r,\th$,
we obtain
\be
\l{3.7}
\c_T =T_{00}\f{\pa}{\pa t}+
\f{K^2-1}{K^2}\left(-T\sqrt{1-b^2}+T' b\right )\f{\pa}{\pa r}
-\f{2T'}{K(\sqrt{1-b^2}-b' )}\f{\pa}{\pa\th}.
\ee
In the near-horizon region, denoting $r=2+x,\,\,\,|x|\ll 1$, we have
$$
K\simeq 1+\sqrt{2x},\,\quad b=b_0 (1-\sqrt{2x}),\qquad b_0 =2C',\qquad \pa K/\pa r \simeq
1/\sqrt{2x}. 
$$
Acting by the generator of supertranslations on the component 
$g_{tt}$ of the metric  (\ref{2.12}), we obtain
\be
\l{3.8}
L_{\c_T}g_{tt} = \f{2}{r^2}\f{K^2-1}{K^2}\left(-T\sqrt{1-b^2}+T' b \right)
.\ee
In the near-horizon region
\be
\l{3.9}
g_{tt}=\f{x}{2}+O(x^2 ).
\ee
To preserve the functional form (\ref{3.9}) of the metric component 
at $x\rightarrow 0$, the 
transformed component (\ref{3.8}) should be of order $O(x)$, or less.
Because $(K^2 -1)/K^2 = O(x^{1/2})$, the  sufficient condition is
\be
\l{3.10}
-T\sqrt{1-b^2}+T' b =O(x^{1/2})
,\ee
and
\be
\l{3.10a}
\f{K^2-1}{K^2}\left(-T\sqrt{1-b^2}+T' b \right)=O(x).
\ee
The action of the term $\c_T^\th\pa_\th$ on the metric components do not change their 
behavior as $x\rightarrow 0$. Let us consider the action of $\c_T^r\pa_r$. When 
acting on $\b{g}_{rr},\,\, \b{g}_{r\th},\,\, \b{g}_{\th\th},\,\,\b{g}_{\p\p}$
the operator $\c_T^r\pa_r$ produces the terms proportional to $V^{-1/2}$. When acting on
the terms $V^{-1}$ and $V^{-1/2}$ the operator  $\c_T^r\pa_r$ produces an extra 
factor $V^{-1}$. However, because of condition (\ref{3.10a}), the total singularity
of each term remains the same or reduces. 

Condition (\ref{3.10}) gives a restriction on supertranslation $T(\th )$
\be
\l{3.11}
-T\sqrt{1-b^2_0}+T' b_0 =0.
\ee
Using (\ref{3.11}) we obtain
\be
\l{3.15}
-T\sqrt{1-b^2}+T' b = -\f{\sqrt{2x}T}{\sqrt{1-b_0^2 }} = O(x^{1/2})
.\ee
To obtain the  relation (\ref{3.15}) we have used (\ref{3.11}).
Eq. (\ref{3.11}) is the first-order equation on $T(\th)$. $T(\th)$ is expressed  
through $C(\th )$ and is determined up to an arbitrary constant.

\section{Asymptotic horizon symmetries of the metric}

In this section we consider general diffeomorpisms which do
not change the form of the metric components in the near-horizon region.
We find transformations of the leading-order parts of the metric components.

At  $x=r-2 \ll 1 $ the metric (\ref{2.12}) takes the form
\ba
\l{4.1}
&{}&ds^2 =(-\k x +O(x^2 ))dt^2 + \left(\f{g_{rr,-1}}{x} +O(x^{-1/2} )\right)dx^2+
2\left(\f{g_{r\th,-1/2}}{x^{1/2}}+O(x^0 \right)dxd\th + \\\nonumber
&{}&+(g_{\th\th,0}+O(x^{1/2})d\th^2 +(g_{\p\p,0}+O(x^{1/2}))d\p^2 .
\ea
The coefficients of expansions of the metric components in series in $x^{1/2}$ are functions
of $\th$. Here $\k$ is introduced to treat the component $g_{tt}$ on the same footing as other
components. In the final expressions $\k$ is set to unity.

We consider transformations generated by vector fields $\c^i$ which do 
not change the form of the metric components in the leading orders.
The vector fields generating the near-horizon transformations have the following structure
\be
\l{4.2}
\c^k = \c^k_0 +x^{1/2}\c^k_{1/2} +x \c^k_{1} +\cdots
\ee
Under the diffeomorphisms generated by vector fields $\c^k $ the metric
components are transformed as
\be
\l{4.3}
L_\chi g_{mn} =\chi^k\pa_k {g}_{mn}+\pa_m \chi^k {g}_{kn} + \pa_n \chi^k  {g}_{km}
.\ee
Transformations preserving the gauge conditions $g_{rt}=g_{\th t}=0$ are
\ba
\l{4.4}
L_{\c}{g}_{rt}=\pa_r \c^t {g}_{tt} +\pa_t\c^r {g}_{rr}+\pa_t \c^\th {g}_{\th r}=0,\\
\l{4.5}
L_{\c}{g}_{\th t}=\pa_\th \c^t	{g}_{tt}+\pa_t\c^r {g}_{r\th}+\pa_t \c^\th {g}_{\th\th}=0.
\ea
Other components of the metric transform as
\ba
\l{4.61}
&{}&L_\c g_{tt}=\c^r \pa_r g_{tt} +\c^\th \pa_\th g_{tt} +2\pa_t\c^t g_{tt} =O(x),\\
\l{4.62}
&{}&L_\c g_{rr}=\c^r \pa_r g_{rr} +\c^\th \pa_\th g_{rr}+2 \pa_r\c^r g_{rr}+
\pa_r\c^\th g_{r\th}=O(1/x),\\
\l{4.63}
&{}&L_\c g_{r\th}= \c^r \pa_r g_{r\th} + \c^\th \pa_\th g_{r\th}+\pa_\th\c^r g_{rr}+
\pa_\th\c^\th g_{r\th}+\pa_r\c^r  g_{r\th}+\pa_r\c^\th  g_{\th\th}=O(x^{-1/2}),\\
\l{4.64}
&{}&L_\c g_{\th\th}= (\c^r \pa_r +\c^\th \pa_\th )g_{\th\th}+2 \pa_\th\c^r g_{r\th} +
2 \pa_\th\c^\th g_{\th\th}=O(x^0 ).
\ea
From the Eq.(\ref{4.62}) it follows that
\be
\c^r_0 = \c^r_{1/2} =0.
\ee
From the Eqs.(\ref{4.4})-(\ref{4.5}) we obtain
\ba
\l{4.7}
\dot{\c}^\th_0 =\dot{\c}^\th_{1/2}  = \dot{\c}^r_{1}=0.
\ea
Here dot is differentiation over $t$.

Transformations of the leading-order parts of the metric components are
\ba
\l{4.16}
&{}&\d\k = \c^r_1\k +\c^\th_0\pa_\th\k +2\dot{\c}^t_0\k ,  \\\nonumber
&{}& \d g_{rr,-1}=\c^r_1 \,g_{rr,-1}+\c^\th_{1/2}\,g_{\th r,-1/2} +\c^\th_0\, \pa_\th g_{rr,-1} ,
 \\\nonumber
&{}&\d g_{\th r,-1/2}=\f{1}{2}\c_1^r \,g_{\th r,-1/2} +\c^\th_0\,g_{\th r,-1/2} +
\pa_\th\c^\th_0 g_{\th r,-1/2} +\f{1}{2}\c^\th_{1/2}g_{\th\th,0} ,\\\nonumber
&{}&\d g_{\th\th ,1/2}= \f{1}{2}\c^r_1\, g_{\th\th ,1/2} + 2\pa_\th\c^r_1\, 
g_{r\th ,-1/2} +\c^\th_0\pa_\th\, g_{\th\th ,1/2} +\c^\th_{1/2}\pa_\th\, g_{\th\th ,0}
+\\\nonumber
&{}& +2\pa_\th\c^\th_0\,g_{\th\th ,1/2} +2 \pa_\th\c^\th_{1/2}\, g_{\th\th	,0}
\\\nonumber
&{}&\d g_{\p\p ,0 }=\c^\th\pa_\th g_{\p\p ,0 }
.\ea
It is seen that the leading-order parts of the metric components are transformed through
the functions
\be
\l{4.17}
\c^r_1(\th),\quad \c^\th_0 (\th),\quad \c^\th_{1/2}(\th)
\ee
and $\dot{\c}^t_0$.

The vector fields in the generator of supertranslations (\ref{3.6}) satisfying
condition (\ref{3.10}) form a subset of the vector fields (\ref{4.2})
\ba
&{}&\c^r =x\c^r_1 = -x\f{T}{\sqrt{1-b_0^2}},\\
&{}& \c^\th =\c^\th_0 =-\f{T'}{\sqrt{1-b_0^2}-b'_0}.
\ea
The Lie brackets of the vector fields in which we retain the components (\ref{4.17}) are
\ba
\l{4.18}
&{}&[\c_{(1)}, \c_{(2)}  ]^t_0 = \c^t_{(12),0} =
\c^t_{(1),0}\stackrel{\leftrightarrow}{\pa}_t\c^t_{(2),0} +
\c^\th_{(1),0}\pa_\th \c^t_{(2),0}-
\c^\th_{(2),0}\pa_\th \c^t_{(1),0}\\ 
&{}&[\c_{(1)}, \c_{(2)} ]^{r}_{1} = \c^r_{(12),1}=
\c^\th_{(1),0}\pa_\th \c^r_{(2),1}-\c^\th_{(2),0}\pa_\th \c^r_{(1),1}\\
&{}&[\c_{(1)}, \c_{(2)} ]^\th_0 = \c^\th_{(12),0}= 
\c^\th_{(1),0}\stackrel{\leftrightarrow}{\pa}_\th\c^\th_{(2),0}\\
&{}&[\c_{(1)}, \c_{(2)} ]^\th_{1/2} = \c^\th_{(1),0}\pa_\th\c^\th_{(2),1/2}-
\c^\th_{(2),0}\pa_\th\c^\th_{(1),1/2}+1/2 (\c^r_{(1),1}\c^\th_{(2),1/2}-
\c^r_{(2),1}\c^\th_{(1),1/2})
\ea
In the next section we use transformations (\ref{4.16}) to calculate 
the surface charge of the asymptotic horizon symmetries.

\section{The surface charge of asymptotic horizon symmetries}

In this section we calculate the variation of the surface charge corresponding to the 
asymptotic Killing vector field $\c^i$. 
The metric variations needed to obtain the variation of the charge were obtained 
in (\ref{4.16}). Details of the calculation are presented in Appendix B. 
In our calculations we used the relations specific to the case of $C$ depending only
on $\th$.

Let us introduce compact  notations for the metric components in (\ref{4.1})
\ba
\l{5.1}
\begin{array}{c} {}\\g_{\m\n}=\\{}\\{}\end{array} 
\left|\begin{array}{cccc}-\k x &0&0&0\\
0 & \b{g}_{rr}/x & \b{g}_{r\th}/\sqrt{x}&0\\
0 & \b{g}_{r\th}/\sqrt{x} & \b{g}_{\th\th} & 0\\
0 & 0 & 0 & \b{g}_{\p\p}\sin^2\th
\end{array}\right|
.\ea
Determinant of the matrix (\ref{5.1}) is
\be
\l{5.3}
|det g|=\k (\b{g}_{rr}\b{g}_{\th\th}-\b{g}_{r\th}^2 )\b{g}_{\p\p}\sin^2\th
=\k \b{g}_{\p\p}\b{g}_{\th\th}\sin^2\th
.\ee
To obtain (\ref{5.3}) we have used the relation
\be
\l{5.4}
\b{g}_{rr}\b{g}_{\th\th}-\b{g}_{r\th}^2 =\b{g}_{\th\th}.
\ee
The inverse metric is
\ba
\l{5.2}
\begin{array}{c} {}\\g^{\m\n}=\\{}\\{}\end{array}
\left|\begin{array}{cccc}(-x\k )^{-1}&0&0&0\\
0&x  & -\sqrt{x}\b{g}_{r\th}/\b{g}_{\th\th} & 0\\
0& -\sqrt{x}\b{g}_{r\th}/\b{g}_{\th\th} & \b{g}_{rr}/\b{g}_{\th\th} & 0\\
0&0&0&(\b{g}_{\p\p}\sin^2\th )^{-1}
\end{array}\right|
\ea
The matrix of metric variations   is 
\ba
\l{5.5}
\begin{array}{c} {}\\h_{\m\n}\equiv
\d g_{\m\n}=\\{}\\{}\end{array}
\left|\begin{array}{cccc}-{\d\k}{x}&0&0&0\\
0& \bar{h}_{rr} /x& \bar{h}_{r\th}/x^{1/2}&0\\
0&\bar{h}_{r\th}/x^{1/2}&\bar{h}_{\th\th}&0\\
0&0&0& \b{h}_{\p\p}\sin^2\th 
\end{array}\right|
.\ea
Variations with the upper indices are defined as 
 ${h}^{\m\n}=
{g}^{\m\r}\d {g}_{\r\la}{g}^{\la\n}$.
 Using relation (\ref{5.4}) again, we find that ${h}^{rr}=0$:
\ba
\l{5.6}
{h}^{rr}=
(g^{rr})^2 \d {g}_{rr}+2{g}^{rr}{g}^{r\th}\d {g}_{r\th}+
({g}^{r\th})^2\d{g}_{\th\th}=
x^2\d \left(\b{g}_{rr}-\f{\b{g}^2_{r\th}}{\b{g}_{\th\th}}\right)=0
.\ea
Taking the trace of variations, we have
\ba
\l{5.7}
&{}&h =
h_{\m\n}g^{\m\n}=\f{\d\k}{\k}+\d \b{g}_{rr}\b{g}^{rr}+2\d \b{g}_{r\th}\b{g}^{r\th}+
\d \b{g}_{\th\th}\b{g}^{\th\th}+\d \b{g}_{\p\p}\b{g}^{\p\p}=\\\nonumber
&{}&=\f{\d\k}{\k} +\f{\d (\b{g}_{rr}\b{g}_{\th\th}- \b{g}^2_{r\th})}{\b{g}_{\th\th}}+
\f{ \d\b{g}_{\p\p}}{\b{g}_{\p\p}}=
\f{\d\k}{\k}+\f{\d \b{g}_{\th\th}}{\b{g}_{\th\th}}+\f{\d \b{g}_{\p\p}}{\b{g}_{\p\p}}
.\ea
Variation of the surface charge is \cite{bar_br,bar_com}
\be
\l{5.8}
\not\d Q_\c (g,h)= \f{1}{16\pi}\int (d^2 x)_{\m\n}F^{\m\n}
,\ee
where $(d^2 x)_{\m\n}=1/4 \e_{\m\n\a\bt} dx^\a \wedge dx^\bt$.
Explicitly
\ba
\l{5.9}
\not\d \hat{Q}_\c (g,h) =\f{1}{4\pi}\int (d^2 x)_{rt}\sqrt{g} \left[\c^r\nabla^t h -
\c^r\nabla_\s h^{t\s} +\c_\s\nabla^r h^{t\s}+\f{1}{2}h\nabla^r\c^t +
\right.\\\nonumber
+\left.\f{1}{2}h^{r\s}(\nabla^t \c_\s -\nabla_\s \c^t) 
-(r\leftrightarrow t) \right].
\ea
Calculating  the integrand  $F^{rt}$, we obtain
\ba
\l{5.10}
&{}&F^{rt}= \f{\c^t_0}{2}\left[ 0\big /-\f{\d\k}{\k} \big/+\f{\d \k}{\k}\big/+
\f{\d \k}{\k}+\f{\d \b{g}_{\th\th}}{\b{g}_{\th\th}}+\f{\d \b{g}_{\p\p}}{\b{g}_{\p\p}}\big/ 
-\f{\d \k}{2\k}\big/-\f{\d \k}{2\k}\right] 
+O(x^{1/2})=\\\nonumber
&{}&=\f{\c^t_0}{2}\left(\f{\d \b{g}_{\th\th}}{\b{g}_{\th\th}}+\f{\d \b{g}_{\p\p}}{\b{g}_{\p\p}}
\right)\bigg|_{x=0} +O(x^{1/2}).
\ea
Here $\big/$ divides contributions $O(x^0 )$  from the consequitive terms in 
(\ref{5.9}). Details of the calculation are presented in Appendix B.

Substituting the above expressions, we obtain the variation of the surface charge 
\ba
\l{5.11}
&{}&\not\d \hat{Q} =\f{1}{16\pi}\int d\p d\th \sin\th {\c_0^t}
(\k \b{g}_{\th\th}\b{g}_{\p\p})^{1/2}\left(\f{\d \b{g}_{\th\th}}{\b{g}_{\th\th}}+
\f{\d \b{g}_{\p\p}}{\b{g}_{\p\p}}\right)=\\\nonumber
&{}&=\f{1}{16\pi}\int d\p d\th \sin\th \c_0^t
\k^{1/2}\f{\d (\b{g}_{\th\th}\b{g}_{\p\p})}{(\b{g}_{\th\th}\b{g}_{\p\p})^{1/2}}.
\ea
Here integration is over the surface $r=2$.
Setting  $\k =1$ in  correspondence with the metric (\ref{2.12})
and integrating the variation (\ref{5.11}), we obtain
\be
\l{5.12} 
\hat{Q} =\f{1}{8\pi} \int d\p d\th \sin\th {\c_0^t} (\b{g}_{\th\th}\b{g}_{\p\p})^{1/2}
+\hat{Q}_0.
\ee
If $\c_0^t$ is independent of $\th$,  the surface charge is proportional 
to the surface of the horizon, i.e.
to the geometric entropy of the black hole.

The Lie bracket of the charges is
\be
\l{5.13}
[\hat{Q}(\c^t_{(1),0}),\hat{Q}(\c^t_{(2),0}) ]= \hat{Q}(\c^t_{(12),0}),
\ee
where $\c^t_{(12),0}$ is defined in(\ref{4.18}).

\section{Conclusion and discussion}

In this note we studied the horizon symmetries of the metric containing 
supertranslation field  
depending  only on spherical angle $\th$.
We transformed the metric obtained in \cite{comp2} to the form with
horizon located at the surface $r=2M$. After reviewing the action of 
supertranslations preserving the form of the metric \cite{comp2}, we
have determined the requirements on the generator of supertranslations
to preserve the near-horizon form of the metric.
It was found that   
to preserve the functional form of the metric
in the near-horizon region, at $r-2M \ll 1$, the $r$-component
of the generator of supertranslations should be of order $O(r-2M)$.
From this requirement it follows  that parameter of supertranslations,
$T(\th )$, must satisfy a condition
$$
-T(\th )\sqrt{1-4{C'(\th )}^2} +2T' (\th) C'(\th) =0
.$$

Next, we studied the general transformations preserving the near-horizon form of 
the metric and 
calculated the charge corresponding to the asymptotic horizon Killing symmetries.
In calculation of the variation of the charge we used the relations 
between the metric components
specific to the case of the supertranslation field depending only on angle $\th$.
In the case of supertranslation field depending only on $\th$
the variation of the charge can be integrated to a charge in a closed form.
 The charge is proportional to the area 
of horizon surface i.e. to the geometric entropy of a black hole.

In papers (a partial list
 \cite{carlip,koga,hotta,padm,donnay41,donnay4,akh,set})  and in many subsequent papers
 the near-horizon symmetries 
were studied for the metrics with the isolated horizon, which
near the horizon have the form 
\be
\l{C}
ds^2=-2\k \r dv^2 +2dvd\r+2\theta_a \r dv dx^a +(\Omega \g_{ab}+\la_{ab}\r )dx^a dx^b
.\ee
The metric is written in the gauge $g_{\r\r}=g_{v\r}=g_{\r a}=0$ with accuracy 
$O(\r^2 )$. Horizon of the metric 
is located at $\r =0$. In the near-horizon region the metric components are 
expanded in power series in $\r$ .
The charge of the asymptotic near-horizon symmetries for the metric (\ref{C})
was obtained in \cite{donnay41,donnay4} in a form
$$
Q=\f{1}{16\pi G}\int dz d\z \sqrt{|\g|}(2 T\k\Omega-y^a\theta_a\Omega )
.$$
Here $T$ is a part of the asymptotic Killing vector $\c^v$.
The volume $\sqrt{\g}\Omega dz d\z$, where $\Omega_{z\z}= \Omega \g_{z\z}$, 
  is an analog of the volume 
$(\b{g}_{\th\th}\b{g}_{\p\p})^{1/2} \sin\th d\th d\p$ in (\ref{5.12}),
and $T$ is an analog of $ \c^t$. It is seen that  structures  of charges in both cases
are similar.

In distinction to the metric (\ref{C}), 
the components of  the metric (\ref{2.12}) are expanded 
in powers of $ x^{1/2}=(r-2M)^{1/2}$. 
In the limit of zero 
supertranslation field the terms with fractional 
powers of  $(r-2M)$ vanish.


\section{Appendix A}
\setcounter{equation}{0}
\renewcommand{\theequation}{A\arabic{equation}}

In this Appendix we consider a solution of the geodesic equations in the 
metric (\ref{2.12}). We find an asymptotic of a solution for null geodesic
in the limit $V(r)\rightarrow 0$.
Following the treatment of \cite{chand}
we start from the Lagrangian corresponding
to the metric (\ref{2.13})
\ba
\l{1.g}
2{\cal{L}}=-V{\dot{t}}^2 +\f{ {\dot{r}}^2 }{V}
\b{g}_{rr}
+2\f{\dot{r}\dot{\th}}{V^{1/2}}
\b{g}_{r\th }
+ {\dot\th}^2 \b{g}_{\th\th}  +
 {\dot\p}^2 \sin^2\th   \b{g}_{\p\p}
.\ea
The derivatives are taken with respect to an affine parameter on geodesic $\tau$.
The function $\p$ is the  cyclic variable which it is possible to set  equal to zero
\cite{chand}.
The Lagrange equations for null geodesics are
\ba\l{1}
&{}&\f{d (V\dt{t})}{d \tau}=0,\\\l{2}
&{}&\f{ \ddot{r}\b{g}_{rr} }{V} -\f{ {\dot{r}}^2 \b{g}_{rr,r} V_{,r} }{2V^2}
+\f{ {\dt{r}}^2 \b{g}_{rr,r} }{2V}+
\f{ \dt{r}\dt{\th}\b{g}_{rr,\th} }{V} +\f{ \ddot{\th}\b{g}_{r\th} }{V^{1/2}}+
\f{ {\dt{\th}}^2 \b{g}_{r\th ,\th} }{V^{1/2}}-\f{ {\dt{\th}}^2 \b{g}_{\th\th,r} }{2} +
\f{ {\dt{t}}^2 V_{,r}}{2}=0,\\\l{3}
&{}& \f{\ddot{r} \b{g}_{r\th} }{V^{1/2}} -\f{ {\dt{r}}^2 \b{g}_{r\th}V_{,r} }{2V^{3/2}}
+\f{ {\dt{r}}^2 \b{g}_{r\th,r}}{V^{1/2}} -\f{ {\dt{r}}^2 \b{g}_{r\th,r}}{V}
+\ddot{\th}\b{g}_{\th\th} +\dot{r}\dot{\th}\b{g}_{\th\th,r} +
\f{{\dt{\th}}^2 \b{g}_{\th\th,\th}}{2}
=0.
\ea
The system of equations admits the first integral
\be\l{4}
-\f{E^2}{2V}+\f{ {\dt{r}}^2 \b{g}_{rr}}{2V} + \f{\dot{r}\dot{\th}\b{g}_{r\th}}{V^{1/2}} +
\f{ {\dt{\th}}^2 \b{g}_{\th\th}}{2}=0.
\ee
We look for solution in the limit $V(r)\rightarrow 0$ in a form
\ba\l{5}
&{}&\dt{t} =\f{E}{V}\\\l{5a}
&{}&\dt{r} =C+C_1 V^{1/2} +...\\\l{6}
&{}&\dt{\th} =\f{A}{V^{1/2}}+A_1 +...
\ea
In the limit $V=0$ we have $r =2M=2,\,\,V_{,r}=2M/r^2 =1/2.$
Substituting the Ansatz in the Lagrange equations and retaining the leading in 
$V\rightarrow 0$
terms, we have
\ba\l{7}
&{}& -\f{C^2 \b{g}_{rr}}{4V^2} +\f{E^2}{4V^2}-\f{AC\b{g}_{r\th}}{4V^2}=0,\\\l{8}
&{}&-\f{C^2 \b{g}_{r\th}}{4V^{3/2}}-\f{AC\b{g}_{\th\th}}{4V^{3/2}}=0.
\ea
From the system (\ref{7})-(\ref{8}) it follows that
\ba\l{9}
&{}&C\b{g}_{r\th} +2A\b{g}_{\th\th}=0,\\
\l{10}
&{}&E^2 -C^2\left(\b{g}_{rr}-\f{\b{g}^2_{r\th}}{\b{g}_{\th\th}}\right)=0
.\ea
Substituting the Ansatz in relation (\ref{4}), we reduce it to
\be\l{11}
V^{-1}[C^2 \b{g}_{rr} -E^2 +2r CA \b{g}_{r\th }
 +r^2  A^2\b{g}_{\th\th}]=0.
\ee
Using the relation (\ref{5.4}),
$$\b{g}_{rr}\b{g}_{\th\th}-\b{g}^2_{r\th} =\b{g}_{\th\th},
$$
from the Eq. (\ref{10}) we find that 
$$C^2 =E^2.
$$
In the Ansatz (\ref{5a})-(\ref{6})  the relation (\ref{11}) is satisfied identicalliy.  
From (\ref{5}) and (\ref{6}) in the main order in $V\rightarrow 0$ we obtain
\be\l{12}
\f{dr}{dt}=\pm |E|V(r).
\ee
From this relation it follows that the surface $r=2M$ is the surface of infinite 
redshift \cite {LL}.

\section{Appendix B}
\setcounter{equation}{0}
\renewcommand{\theequation}{B\arabic{equation}}

In this Appendix we present some detailes of calculation of the variation of the
 surface charge.
The integrand in (\ref{5.11}) is
\ba
F^{rt}=\left[\c^r\nabla^t h -
\c^r\nabla_\s h^{t\s} +\c_\s\nabla^r h^{t\s}+\f{1}{2}h\nabla^r\c^t +
\f{1}{2}h^{r\s}(\nabla^t \c_\s -\nabla_\s \c^t) 
-(r\leftrightarrow t)\right].
\ea
The first term is
\be
\c^r\nabla^t h-\c^t\nabla^r h =-\c^t\left(x\pa_x -
\f{x^{1/2}\b{g}_{r\th}}{\b{g}_{\th\th}}\pa_\th \right)h
=O(x^{1/2}).
\ee
In the limit $x=0$ its conribution is zero.
The second term is
\be
-\c^r\nabla_\s h^{t\s} + \c^t\nabla_\s h^{r\s}=\c^t\nabla_\s h^{r\s}+O(x^{1/2} )=
-\f{\c^t}{2}\f{\d\k}{\k}+     O(x^{1/2}).
\ee
The third term reduces to
\ba
&{}&\c_\s \nabla^r h^{t\s}-\c_\s \nabla^t h^{r\s}=\c_t (g^{r\th}\nabla_\th h^{tt}
 -g^{tt}\nabla_t h^{rt} )+     O(x^{1/2})= \\\nonumber
&{}&=-\c^t \nabla_t h^{rt}+O(x^{1/2}) =\f{\c^t}{2}\f{\d\k}{\k} +O(x^{1/2})
\ea
The fourth term is
\be
\f{h}{2}(\nabla^r\c^t -\nabla^t\c^r )=\f{h}{2}\left((g^{rr}\nabla_r -
g^{r\th}\nabla_\th )\c^t- g^{tt}\nabla_t\c^r \right) =\f{\c^t}{2}\left(\f{\d\k}{\k}
+\f{\d\b{g}_{\th\th}}{\b{g}_{\th\th}}+\f{\d \b{g}_{\p\p}}{\b{g}_{\p\p}}\right)+O(x^{1/2}).
\ee
The fifth term  yields
\ba
&{}&
\f{1}{2}(h^{r\s}\nabla^t\c_\s -h^{t\s}\nabla^r\c_\s)=\f{1}{2}(h^{rr}\nabla^t\c_r +
h^{r\th}\nabla^t\c_\th - h^{tt}\nabla^r\c_t)=
\\\nonumber
&{}&=\left[\nabla_t\c^r\, g^{tt}\left(h^{rr}g_{rr}+h^{r\th}g_{r\th}\right)+
\nabla_t\c^\th \, g^{tt}\left(h^{rr}g_{r\th}+h^{r\th}g_{\th\th}\right)- \right.   \\\nonumber
&{}&\left. -h^{tt}g_{tt}\left(g^{rr}\nabla_r\c^t + g^{r\th}\nabla_\th\c^t\right)\right]+ O(x^{1/2})=
-\f{\c^t}{2}\f{\d\k}{2\k} +O(x^{1/2}).
\ea
The sixth term is
\ba
&{}&-\f{1}{2}(h^{r\s}\nabla_\s\c^t +h^{t\s}\nabla_r\c^\s )=
\f{1}{2}[h^{tt}\nabla_t\c^r -h^{rr}\nabla_r\c^t -h^{r\th}\nabla_\th\c^t ]+  
   O(x^{1/2})=\\\nonumber
&{}&=-\f{\c^t}{2}\f{\d\k}{2\k} + O(x^{1/2} ).
\ea

\vspace*{1cm}

{\large\bf Acknowledgments}

I thank Misha Smoliakov and Igor Volobuev for interesting discussion.

This work was partially supported by the Ministry of Science and Education of
Russian Federation under the project 01201255504.

$\sin\theta(\bar{g}_{\theta\theta}\bar{g}_{\varphi\varphi} )^{1/2}d\theta d\varphi$
$\Omega_{z\bar{z}}=\Omega\gamma_{z\bar{z}}$

\begin{thebibliography}{99}


\bibitem{bondi}
H. Bondi, M. G. J.van der Burg and A. W. K. Metzner, {\it Gravitational vaves
in general relativity 7}, Proc. Roy. Soc. Lond. {\bf A269}, 21 (1962).
\bibitem{sachs}
R. K. Sachs, {\it Gravitational vaves
in general relativity 8. Waves in asymptotically flat space-time.}
 Proc. Roy. Soc. Lond. {\bf A270}, 103 (1962).
\bibitem{strom1}
A. Strominger, {\it Lectures on the Infrared Structure of Gravity and Gauge 
Theories},
 arXiv:1703.05448.
\bibitem{bar3}
G. Barnich  and C. Troessaert, {\it Aspects of the BMS/CFT correspondence},
JHEP 05, 062 (2010), arXiv:1001.1541.
\bibitem{bar2}
G. Barnich  and C. Troessaert, {\it Symmetries of asymptotically flat
4 dimensional spacetimes at null infinity revisited},
Phys. Rev. Lett. {\bf B105}, 111103 (2010), arXiv:gr-qc/0909.2617.
\bibitem{bar4}
G. Barnich  and C. Troessaert, {BMS charge algebra},
JHEP 12, 105 (2011) , arXiv:1106.0213.
\bibitem{charge}
E. E. Flanagan and D. A. Nichols,
{\it Conserved charges of the extended Bondi-Metzner-Sachs algebra},
        Phys. Rev. {\bf D95},  044002 (2017), arXiv:1510.03386.
\bibitem{comp1}
G. Compere and J. Long, {\it Vacua of the gravitational field}, 
JHEP 07, 137 (2016),
arXiv:1601.04958.

\bibitem{comp2}
G. Compere and J. Long, {\it Classical static final state of collapse with
supertranslation memory},
Class. Quant. Grav. {\bf 33},  195001 (2016), arXiv:1602.05197.

\bibitem{fad}
P. P. Kulish and L. D. Faddeev, {\it Asymptotic conditions and 
infrared divergencies in
quantum electrodynamics}, Theor. Math. Phys. {\bf 4}, 745 (1970).

\bibitem{carney}
D. Carney, L. Chaurette, D. Neuenfeld and  G.Semenoff,
{\it  Infrared quantum information},
 	Phys. Rev. Lett. {\bf 119} (2017) 180502, arXiv:1706.03782.
\bibitem{strom4}
A. Strominger,
{\it Black Hole Information Revisited}, arXiv: 1706.07143.

\bibitem{carter}
B. Carter, { \it Axisymmetric black hole has only two degrees of freedom},
Phys. Rev. Lett. {\bf 26}, 331 (1971).
\bibitem{robinson}
D. C. Robinson, {\it Uniqueness of the Kerr black hole}, Phys. Rev. Lett.
 {\bf 34}, 905 (1975).

\bibitem{hawk3}
S. W. Hawking, M. J. Perry and A. Strominger,
{\it Soft Hair on Black Holes},
Phys. Rev. Lett. {\bf 116}, 231301 (2016), arXiv:1601.00921.

\bibitem{hawk31}
S.W. Hawking, M.J. Perry and A. Strominger,
{\it Superrotation Charge and Supertranslation Hair on Black Holes},
JHEP 05, 161 (2017), arXiv:1611.09175.

\bibitem{bousso1}
R. Bousso and M. Porrati,
{\it Soft Hair as a Soft Wig },  arXiv:1706.00436.

\bibitem{bousso2}
R. Bousso and M. Porrati,
{\it Observable Supertranslations}, Phys. Rev. {\bf D96}, 103512 (2017),
 arXiv:1706.08503.

\bibitem{booth}
I. Booth, {\it Spacetime near isolated and dynamical trapping horizons}, 
 	Phys. Rev. {\bf D87},  024008 (2013), arXiv:1207.6955.
\bibitem{wald}
H. Friedrich, I. Racz and R. M. Wald,
{ On the Rigidity Theorem for Spacetimes with a Stationary
Event Horizon or a Compact Cauchy Horizon},
Commun. Math. Phys.{\bf  204}, 691 (1999),
arXiv:gr-qc/9811021.

\bibitem{carlip}
S. Carlip,
{\it Entropy from Conformal Field Theory at Killing Horizons},
 	Class. Quant. Grav.{\bf 16}, 3327 (1999), arXiv:gr-qc/9906126.
\bibitem{koga}
J.-i. Koga, {\it Asymptotic symmetries on Killing horizons},
 Phys. Rev. {\bf D64},  124012 (2001), arXiv:gr-qc/0107096.

\bibitem{hotta}
M. Hotta, K. Sasaki and T. Sasaki,
{\it Diffeomorphism on Horizon as an Asymptotic Isometry of Schwarzschild Black Hole},
 	Class. Quant. Grav. {\bf 18}, 1823 (2001), arXiv:gr-qc/0011043.

\bibitem{padm}
B. R. Majhi and T. Padmanabhan,
{\it Noether Current, Horizon Virasoro Algebra and Entropy},
 	Phys. Rev. {\bf D 85},  084040 (2012), arXiv:1111.1809.

\bibitem{donnay41}
L. Donnay, G. Giribet, H. A. Gonzalez, and M. Pino,
{\it  Supertranslations and Superrotations at the Black Hole Horizon},
Phys. Rev. Lett. {\bf 116},  091101 (2016),  arXiv:1511.08687.

\bibitem{donnay4}
L. Donnay, G. Giribet, H. A. Gonzalez, and M. Pino,
{\it Extended symmetries at the black hole horizon},
JHEP 09, 100 (2016), arXiv:1607.05703.

\bibitem{akh}
E. T. Akhmedov and M. Godazgar,
{\it  Symmetries at the black hole horizon},
Phys. Rev. { \bf D 96},   104025 (2017), arXiv:1707.05517.

\bibitem{set}
M. R. Setare and H. Adami,
{\it BMS type symmetries at null-infinity and near horizon
of non-extermal black holes},  	Eur. Phys. J. {\bf C76}, 687 (2016), 
 arXiv:1609.05736.

\bibitem{iofa}
M. Z. Iofa, {\it Thermal Hawking radiation of black hole with 
supertranslation field},
JHEP 01, 137 (2018),  arXiv:1708.09169.

\bibitem{LL}
L. D. Landau and E. M. Lifschiz {\it Classical theory of fields}, Moscow (1990).

\bibitem{bar_br}
G. Barnich and F. Brandt,
{\it Covariant theory of asymptotic symmetries, conservation laws and central charges},
 	Nucl. Phys. {\bf B633}, 3 (2002), arXiv:hep-th/0111246.

\bibitem{bar_com}
G. Barnich and G. Compere,
{\it  Surface charge algebra in gauge theories and thermodynamic integrability},
 	J. Math. Phys. {\bf 49},  042901 (2008), arXiv:0708.2378.

\bibitem{chand}
S. Chandrasekhar {\it The mathematical theory of black holes}, 
Oxford University Press, (1983).


\end{thebibliography}
\end{document}